# Solvable Schrödinger Equations of Shape Invariant Potentials with Superpotential $W(x, A, B) = A \tanh 3px - B \coth px$

Jamal Benbourenane

Abu Dhabi, UAE

We propose a new, exactly solvable Schrödinger equation. The potential with the shape invariance property is given by
$$V = -Bp\, csch[px]^2 - 9p(B+p)sech[3px]^2 + (B\coth[px] - 3(B+p)\tanh[3px])^2.$$
The associated superpotential is $W(x, A, B) = A \tanh 3px - B \coth px$.

We derive entirely the exact solutions of this family of Schrödinger equations with the eigenvalues (energies) given by the formula
$$E_n^{(-)} = (A-B)^2 - (A-B-4np)^2,$$
and the corresponding eigenfunctions determined exactly and in closed form.

Schrödinger equations, and Sturm-Liouville equations in general, are challenging to solve in closed form, and only a few of them are known. Therefore, in a strict mathematical sense, discovering new solvable equations is essential in understanding the eluded solutions' underpinnings. This result has potential applications in nuclear physics and chemistry, and other fields of science.

## 1   Introduction

Schrödinger equation was developed (1926) [36] by Erwin Schrödinger to study the atomic spectra. It has many science applications, including nuclear physics, molecular chemistry, nanostructures, quantum processing, quantum biology, quantum finance, etc. It is pivotal for all these fields of science. In the strict mathematical sense of exactness, exact solutions are the ultimate goal when studying these solutions' properties.

Many authors contributed immensely to the analytical solutions of the Schrodinger equation. This article focuses on exact solutions given explicitly and in the closed forms of the one-dimensional Shrödinger equation. It is common knowledge that the only solvable potentials available in the literature are very limited [37] [30] [17] [33] [32] [35] [4] [6]. All these potentials, surprisingly, have the shape invariance property.

We will use Supersymmetry (SUSY) for evaluating the eigenvalues and eigenfunctions of the Schrödinger equation. This technique was introduced by Nicolai and Witten in non-relativistic quantum mechanics [31] [40]. It is based on the factorization method that dates back to Bernoulli, Cauchy, and the best-known work of Darboux [13] in 1882. It took almost a century for Darboux's seminal paper for recognition.

In this paper, we determine the exact spectrum and the same corresponding eigenfunctions of the potential in the form
$$V_-(x, A, B) = (A-B)^2 + \frac{1}{4}B(B-2p)csch^2 px - \frac{1}{4}(2A-B)(2A-B+2p)sech^2 px$$

with superpotential given by
$$W(x, A, B, p) = A\tanh 3px - B\coth px.$$

In the next section, we introduce the concept of SUSY. Section 3 is devoted to the shape invariance property that allows generating exact solutions. In section 4, we will completely solve the new potential using the SUSY method. This potential satisfies the shape invariance property. We, therefore, can evaluate its eigenvalues and eigenfunctions in closed form. We conclude by summarizing our discovery and its implications in section 5.

## 2   Supersymmetry

Given a potential $V_-(x)$, we seek to build a potential partner $V_+(x)$, where these two potentials have the same energy eigenvalues, except for the ground state.

These potential partners are defined by
$$V_\pm(x) = W^2(x) \pm W'(x) \tag{1}$$

where $W(x)$ is called the superpotential.

The associated Hamiltonian to these potential partners are given by
$$H_- = A^\dagger A, \qquad H_+ = AA^\dagger \tag{2}$$

where
$$A^\dagger = -\frac{d}{dx} + W(x), \qquad A = \frac{d}{dx} + W(x) \tag{3}$$

The one dimensional time-independent Schrödinger equation with eigenstate $E$ and potential $V(x)$ is defined by
$$\left(-\frac{d^2}{dx^2} + V(x)\right)\Psi = E\Psi \tag{4}$$

and the two Hamiltonians associated with the Schrödinger equation are written in the form
$$H_- = -\frac{d^2}{dx^2} + V_-(x), \quad H_+ = -\frac{d^2}{dx^2} + V_+(x), \tag{5}$$

These two Hamiltonians (5) have their eigenvalues and their eigenfunctions intertwined. That is, if $E_0^{(-)}$ is an eigenstate with eigenfunction $\Psi_0^{(-)}$, then the Hamiltonian $H_+$ will have the same eigenstate $E_0^{(-)}$ and its eigenfunction is given by $A\Psi_0^{(-)}$, and vice-versa if we change $H_+$ by $H_-$.

The two Hamiltonians are both positive semi-definite operators, so their energies are greater than or equal to zero.

For the Hamiltonian $H_-$, we have
$$H_-\Psi_0^{(-)} = A^\dagger A \Psi_0^{(-)} = E_0^{(-)}\Psi_0^{(-)} \tag{6}$$

So, by multiplying on the left side of the equation (6) by the operator $A$, we obtain

$$AA^\dagger A\Psi_0^{(-)} = E_0^{(-)}(A\Psi_0^{(-)}),$$

so that

$$H_+(A\Psi_0^{(-)}) = E_0^{(-)}(A\Psi_0^{(-)}).$$

Similarly, for $H_+$

$$H_+\Psi_0^{(+)} = AA^\dagger\Psi_0^{(+)} = E_0^{(+)}\Psi_0^{(+)} \qquad (7)$$

and multiplying the left side of (7) by $A^\dagger$ we obtain

$$H_-(A^\dagger\Psi_0^{(+)}) = E_0^{(+)}(A^\dagger\Psi_0^{(+)}). \qquad (8)$$

The two Hamiltonians' eigenfunctions and their exact relationships will depend on the quantity $A\Psi_0^{(-)}$. If $E_0^{(-)}$ is zero (nonzero), it means an unbroken (broken) supersymmetric system [38] [39].

Thus, here we will consider only the case of unbroken supersymmetry, where $A\Psi_0^{(-)} = 0$. In this case, this state has no SUSY partner since the ground state wavefunction of $H_-$ is annihilated by the operator $A$, and in which case $E_0^{(-)} = 0$.

It is then clear that the two Hamiltonians $H_-$ and $H_+$'s eigenstates and eigenfunctions are related by (for $n = 0,1,2,\ldots$)

$$E_0^{(-)} = 0, \quad E_n^{(+)} = E_{n+1}^{(-)}, \qquad (9)$$

$$\Psi_n^{(+)} = \left(E_{n+1}^{(-)}\right)^{-1/2} A\Psi_{n+1}^{(-)} \qquad (10)$$

$$\Psi_{n+1}^{(-)} = \left(E_n^{(+)}\right)^{-1/2} A^\dagger\Psi_n^{(+)}. \qquad (11)$$

Note that $\Psi_0^{(-)}$ is a solution to the Schrödinger Eq. (4) with potential $V$, that is,

$$\Psi_0 = \Psi_0^{(-)} = N\ e^{-\int W(x)dx}$$

where $N$ is the normalizing constant.

The bound state wavefunctions must converge to zero at the two ends of its interval domain. Therefore the two statements $A\Psi_0^{(-)} = 0$ and $A^\dagger\Psi_0^{(+)} = 0$ cannot be satisfied simultaneously, and so only one of the two ground-state energies, $E_0^{(-)}$ and $E_0^{(+)}$ can be zero. In contrast, the other bound state energy will be positive. So, we will consider $W$ such that $\Psi_0$ is normalized satisfying the positive semi-definite condition:

$$E_0^{(-)} = 0,\ E_1^{(-)} = E_0^{(+)} > 0. \qquad (12)$$

## 3  Shape Invariance

In our quest to construct the superpotential $W$ in (1) by solving the Riccati equation, we have looked at the solvable Schrödinger equations among the thirteen known solvable models, and we have investigated the pattern that governs these potentials $V(x)$. These potentials are having in common a particularly interesting geometric property in the shape of the potential partners, as it can be seen in the potentials which was mentioned in the following references [4] [5] [6] .

We will define potentials to be shape invariant if their dependence on $x$ is similar, and they only differ on some parameters appearing in their expressions. The following relation describes this similarity:

$$V_-(x, a_1) + h(a_1) = V_+(x, a_0) + h(a_0) \qquad (13)$$

where the parameter $a_1$ depends on $a_0$, i.e. $a_1 = f(a_0)$, and then, $a_2 = f(a_1) = f^2(a_0)$, and by recurrence $a_k = f^k(a_0)$ where $a_0 \in \mathbb{R}^m$, and $f: \mathbb{R}^m \to \mathbb{R}^m$, is called a parameter change function, see [1] [8] [10] [12] [19].

With the shape invariance property, the superpotentials have a similar form except in their parameters. Their shapes differ vertically by a nonzero constant.

The condition of shape invariance (13) also proved to be another significant hurdle in finding its solutions. Since only the already known solvable potentials, namely, harmonic oscillator, Coulomb, 3D-oscillator, Morse, Rosen-Morse I &II, Eckart, Scarf I&II, Pőschl-Teller I&II, have been shown to satisfy this condition. Recently, we introduced three more potentials that fulfil the invariance shape property [4] [6]. We derived their exact eigenvalues and eigenfunctions.

More precisely, we aim to determine all the bound state energies and the expression of their wavefunctions.

We will consider the potential $V(x, a)$ defined in the Schrödinger equation (4) as a shape-invariant potential. Therefore, the two potential partners $V_-(x, a_1)$ and $V_+(x, a_0)$ have the same dependence on $x$, up to the change in their parameters, and their Hamiltonians $H_-(x, a_1)$ and $H_+(x, a_0)$ differ only by a vertical shift is given by $C(a_0) = h(a_1) - h(a_0)$,

$$V_+(x, a_0) = V_-(x, a_1) + C(a_0) \qquad (14)$$

where the potential partners are defined by

$$V_-(x, a_1) = W^2(x, a_1) - W'(x, a_1), \qquad (15)$$
$$V_+(x, a_0) = W^2(x, a_0) + W'(x, a_0) \qquad (16)$$

The zero-ground state wavefunction is given by

$$\Psi_0^{(-)}(x, a_0) \propto e^{-\int W(x, a_0) dx}. \qquad (17)$$

By differentiating twice, we can show that this wavefunction satisfies the Schrödinger equation

$$-\Psi_0''(x) + V(x)\Psi_0(x) = 0 \qquad (18)$$

with potential $V(x) = V_-(x, a_0)$ and associated eigenvalue $E_0^{(-)} = 0$.

The first excited state of $H_-(x, a_1)$ is $\Psi_1^{(-)}$, where we omit the normalization constant,

$$\Psi_1^{(-)}(x, a_0) = A^\dagger(x, a_0)\Psi_0^{(+)}(x, a_0) = A^\dagger(x, a_0)\Psi_0^{(-)}(x, a_1). \qquad (19)$$

The eigenvalue associated with this Hamiltonian has the following expression
$$E_1^{(-)} = C(a_0) = h(a_1) - h(a_0) \tag{20}$$

The two Hamiltonians $H_+$ and $H$'s have the same eigenvalues except for the additional zero-energy eigenvalue of the lower ladder Hamiltonian $H_-$. These values are related by
$$E_0^{(-)} = 0, \quad E_{n+1}^{(-)} = E_n^{(+)}, \tag{21}$$
$$\Psi_n^{(+)} \propto A\Psi_{n+1}^{(-)}, A^\dagger \Psi_n^{(+)} \propto \Psi_{n+1}^{(-)}, n = 0,1,2,\ldots \tag{22}$$
where we have iterated this procedure to construct a hierarchy of Hamiltonians

$$H_\pm^{(n)} = -\frac{d^2}{dx^2} + V_\pm(x, a_n) + \sum_{k=0}^{n-1} C(a_k) \tag{23}$$

and then derive the $n^{th}$ excited eigenfunction and eigenvalues by
$$\Psi_n^{(-)}(x, a_0) \propto A^\dagger(x, a_0) A^\dagger(x, a_1) \ldots A^\dagger(x, a_n) \Psi_0^{(-)}(x, a_n) \tag{24}$$

$$\begin{aligned}
E_0^{(-)} &= 0, \\
E_n^{(-)} &= \sum_{k=0}^{n-1} C(a_k) = \sum_{k=0}^{n-1} h(a_{k+1}) - h(a_k) \\
&= h(a_n) - h(a_0), \text{ for } n \geq 1.
\end{aligned} \tag{25}$$

where $a_k = f(f(\ldots f(a_0))) = f^k(a_0), k = 0,1,2,\ldots, n-1$.

Therefore, by knowing the superpotential, not only we know the potential but also its ground state. From the algorithm above, the whole spectrum of the Hamiltonian $H_-$ ($H_+$) can be derived by the supersymmetry quantum mechanics method.

## 4  Exactly solvable potentials with superpotential $W(x, A, B) = A\tanh 3px - B\coth px$

We consider the superpotential
$$W(x, A, B) = A\tanh 3px - B\coth px \tag{26}$$
with its potential partners $V_- = V_-(x, A_1, B_1)$ and $V_+ = V_+(x, A_0, B_0)$ satisfying the shape invariance condition (13), we obtain

$$V_- = (-2B_1 + 3p)^2 - \frac{12B_1(B_1 - p)}{(1 - 2\cosh[2px])^2} - \frac{8B_1(B_1 - p)}{-1 + 2\cosh[2px]} + 4B_1(B_1 - p)csch[2px]^2$$
$$V_+ = (2B_0 + 3p)^2 - \frac{12B_0(B_0 + p)}{(1 - 2\cosh[2px])^2} - \frac{8B_0(B_0 + p)}{-1 + 2\cosh[2px]} + 4B_0(B_0 + p)csch[2px]^2$$

which can be written in a more simplified form by using the sequence defined in (28) (29), and which exhibit the shape invariance property, as the two potentials look similar except in the parameters appearing in them. The main potential in (??) of the Schrödinger equation $V(x)$ is defined for $x > 0$ by $V(x) = V_-(x, A, B)$ and can be written in the form
$$V(x) = -Bpcsch[px]^2 - 9p(B + p)sech[3px]^2 + (B\coth[px] - 3(B + p)\tanh[3px])^2 \tag{27}$$

The sequence $a_k = (A_k, B_k)$ of the shape invariance property is given as follows:
$$A_0 = 3(B_0 + p), \tag{28}$$
$$A_k = A_0 - 3kp,$$
$$B_k = B_0 + kp, \quad k = 0,1,2,.. \tag{29}$$

If we assume $0 < p < B$, this potential satisfies the positive semi-definite condition (12). In fact, it has a negative minimum value at the positive real number $x_0 = \ln t$, where $t$ is a real root of the polynomial equation

$$(B^2 p + 2Bp^2) + (-2B^2 p - Bp^2)t^2 + (9B^2 p + 36Bp^2 + 27p^3)t^4 +$$
$$(-36B^2 p - 114Bp^2 - 81p^3)t^6 + (42B^2 p + 126Bp^2 + 81p^3)t^8 +$$
$$(-36B^2 p - 90Bp^2 - 54p^3)t^{10} + (42B^2 p + 126Bp^2 + 81p^3)t^{12} +$$
$$(-36B^2 p - 114Bp^2 - 81p^3)t^{14} + (9B^2 p + 36Bp^2 + 27p^3)t^{16} +$$
$$(-2B^2 p - Bp^2)t^{18} + (B^2 p + 2Bp^2)t^{20}$$
$$= 0 \tag{30}$$

The first energy level of the potential partner $V_-$ is given by (20),
$$E_1^{(-)} = V_+(x, A_0, B_0) - V_-(x, A_1, B_1) \tag{31}$$
$$= (A_0 - B_0)^2 - (A_1 - B_1)^2$$
$$= (A_0 - B_0)^2 - (A_0 - B_0 - 4p)^2$$
$$= (2B_0 + 3p)^2 - (2B_0 - p)^2$$
$$= 8p(2B_0 + p)$$

Using the formula (25), the $n^{th}$ discrete energy is given by

$$E_n^{(-)} = (A_0 - B_0)^2 - (A_n - B_n)^2, \tag{32}$$
$$= (A_0 - B_0)^2 - (A_0 - B_0 - 4np)^2$$
$$= (2B_0 + 3p)^2 - (2B_0 - (4n - 3)p)^2$$

For $n \geq 0$, this energy can be written as
$$E_n^{(-)} = 8np(2B_0 + 3p - 2np). \tag{33}$$

We also observe that the un-normalizable ground wavefunction expression, obtained using (17) is given by
$$\Psi_0(x, A_0, B_0) = \cosh 3px^{-\frac{B_0+p}{p}} \sinh px^{\frac{B_0}{p}}. \tag{34}$$

The first excited state wavefunction is
$$\Psi_1(x, A_0, B_0) = \left(-\frac{d}{dx} + W(x, A_0, B_0)\right) \Psi_0(x, A_1, B_1) \tag{35}$$
$$= (2B_0 + p)\cosh 3px^{-\frac{B_0+p}{p}} (-2\cosh 2px + \cosh 4px) \sinh px^{\frac{B_0}{p}}$$

and the other wavefunctions are obtained using the recursive formula

$$\Psi_n(x, A_k, B_k) = (-\frac{d}{dx} + W(x, A_k, B_k))\Psi_{n-1}(x, A_{k+1}, B_{k+1}), \tag{36}$$

for $k = 0,..,n-1$, $n = 1,2,...$

The bound state $n$ for the potential $V$ needs to satisfy the physical constraint on the energies $E_n$ (??): for all $n \geq 1$

$$0 \leq E_{n-1}^{(-)} < E_n^{(-)} \tag{37}$$

We can show, under this condition, and if $0 < p < B_0$, then

$$1 \leq n < \frac{2B_0 + 5p}{4p}. \tag{38}$$

The eigenfunctions $\Psi_n(x)$ satisfy the following property:

$$\Psi_n(x) \propto \cosh 3px^{-\frac{B_0+p}{p}} \sinh px^{\frac{B_0}{p}} P_n(x) \tag{39}$$

where $P_n$ is a real hyperbolic polynomial of the form

$$P_n(x) = \sum_{k=0}^{2n} \alpha_k \cosh(2kpx), \tag{40}$$

and where the coefficients $\alpha_k$ are real constants that depend on $B_0$ and $p$, and $k$, with $\alpha_n \neq 0$.

The other physical constraint is related to the normalizing of the eigenstates over the positive real axis. At infinity, the limit should be equal to zero; therefore, from equation (40) we have

$$\lim_{x \to \pm\infty} \Psi_n(x) \propto \lim_{x \to \pm\infty} e^{(-3(B_0+p)+4np+B_0)x}. \tag{41}$$

This constraint implies that $-2B_0 + (4n-3)p < 0$. This inequality gives the least upper bound compared to (38). This implies that the maximum number of bound states $n_{\max}$ is attained when

$$n_{\max} = \begin{cases} \left\lfloor \frac{2B_0+3p}{4p} \right\rfloor & if \quad \frac{2B_0+3p}{4p} \text{ is not an } integer \\ \left\lfloor \frac{2B_0+3p}{4p} \right\rfloor - 1 & if \quad \frac{2B_0+3p}{4p} \text{ is an } integer \end{cases} \tag{42}$$

where here the notation $\lfloor . \rfloor$, represents the integer part of the argument.

In figure 1, we display the plot of the potential $V$ with its discrete energies and eigenfunctions when the parameters $B_0 = 7$, $p = 0.5$, and in this case, according to (??) the maximum number of bound states is $n_{\max} = 7$. All energy discrete state levels are below the asymptotic line $v = (2B_0 + 3p)^2$ as can be seen from (32).

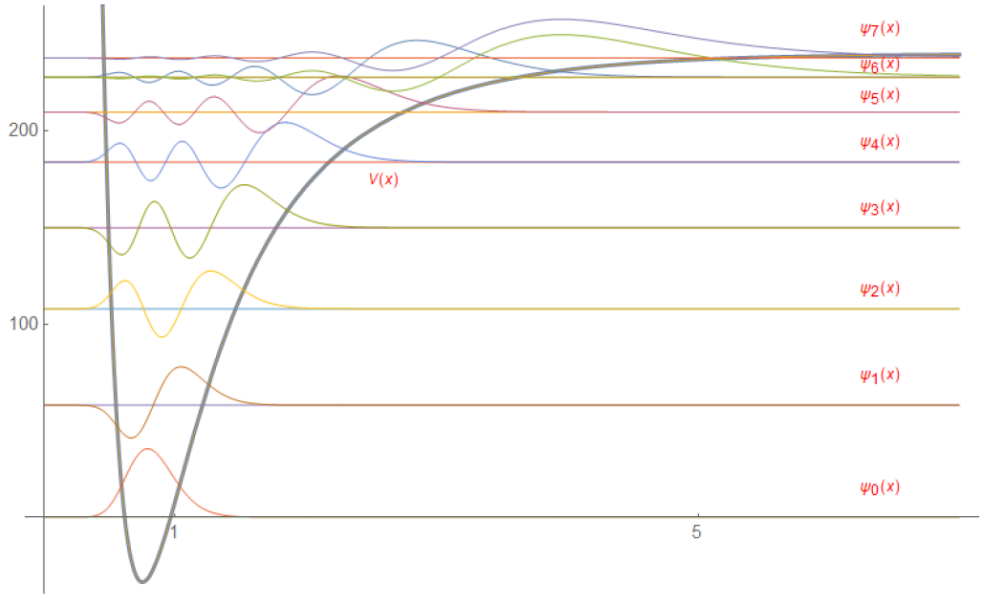

Figure 1. Potential V[x] and the only allowed energy state levels $E_1,\ldots,E_7$ for the parameters B0=7, p=0.5

## 5  Conclusion

We have introduced a new family of exactly solvable Schrödinger equations
$$V(x) = -Bp\,csch[px]^2 - 9p(B+p)sech[3px]^2 + (B\coth[px] - 3(B+p)\tanh[3px])^2,$$
having finite discrete bound states. The energy of the associated partner potential $V_-$ is given exactly in the form
$$E_n^{(-)} = (A-B)^2 - (A-B-4np)^2,$$
and their corresponding eigenfunctions are given recursively and in closed form.

The superpotential associated with this potential is given by
$$W(x,A,B) = A\tanh 3px - B\coth px.$$

This potential has a singularity at zero and a finite number of bound states. We can adjust the number of the discrete bound states, the well's depth and width by varying the parameters. It has applications in different physics and chemistry fields and will be an addition to the list of solvable Schrödinger equations and Sturm-Liouville equations.

## References

[1]  Bagchi, B.K. Supersymmetry in quantum and classical mechanics, Chapman&Hall, (2001).

[2]     Barclay, D. T, Dutt, R, Gangopadhyaya, A,   Khare, A, Pagnamenta, A. and Sukhatme, U. New Exactly Solvable Hamiltonians: Shape Invariance and Self-Similarity, Physical Review A, 48, 4, (1993) pp. 2786-2797. doi:10.1103/PhysRevA.48.2786

[3]     Balantekin, A.B. Algebraic approach to shape invariance, Phys. Rev A., 57, 6 (1998) pp. 4188–4191. 10.1103/PhysRevA.57.4188

[4]     Benbourenane, J., Benbourenane, M,, Eleuch, H. (2021)," Solvable Schrodinger Equations of Shape Invariant Potentials with superpotential  $W(x, A, B) = A\ \tanh(px) + B\ \tanh(6px)$  ". Arxiv.

[5]     Benbourenane, J., Benbourenane, M,, Eleuch, H. (2020)," Exactly Solvable New Classes of Symmetric Triple-Well Potentials. Arxiv.

[6]     Benbourenane, J., Eleuch, H. Exactly Solvable New Classes of Potentials with Finite Discrete Energies. Res. Phys. 17, 103034 (2020)

[7]     Bougie, J, Gangopadhyaya, A, Mallow, J.V. Rasinariu, C. Generation of a novel exactly solvable potential. Phys. Lett. A, 379, (2015) pp. 2180-2183. doi:10.1016/j.physleta.2015.06.058

[8]     Bougie, J, Gangopadhyaya, A, Mallow, J.V. Rasinariu, C. Supersymmetric quantum mechanics and solvable models, Symmetry, 4 (2012) pp 452-473.

[9]     Chuan, C. Exactly solvable potentials and the concept of shape invariance, J. Phys.: Math. Gen. 24 (1991) L1165-L1174.

[10]     Cooper, F., Ginocchio, J. N., and Khare, A. Relationship between Supersymmetry and Solvable Potentials, Physical Review D, 36, 8, (1987) pp. 2458-2473.

[11]     Cooper, F., Khare, A. and Sukhatme, U.   Supersymmetry and Quantum Mechanics, Reports, 251, 5-6, (1995) pp. 267-385. doi:10.1016/0370-1573(94)00080-M

[12]     Dabrowska, J. W., Khare, A. and Sukhatme, U.P.    Explicit Wavefunctions for Shape-Invariant Potentials by Operator Techniques, Journal of Physics A: Mathematical and

General, 21, 4 (1988) pp. L195-L200. doi:10.1088/0305-4470/21/4/002

[13]     Darboux, G. Sur une proposition relative aux equatios lineaires. Compt. Rend. Acad. Sci. 94, 1456 (1882).

[14]     Dutt, R., Khare, A., Sukhatme, U.P. Supersymmetry, shape invariance, and exactly solvable potentials, Am. J. Phys., 56,(1988) pp. 163-168.

[15]     Dutt, R, Khare, A, and Sukhatme, U.P.   Exactness of Supersymmetric WKB Spectra for Shape-Invariant Potentials, Phys. Lett. B, 181, 3-4,(1986) pp. 295-298. doi:10.1016/0370-2693(86)90049-3

[16]     Dutt, R, Gangopadhyaya, A, Khare, A, Pagnamenta, A, Sukhatme, U. Solvable quantum mechanical examples of broken supersymmetry, Phys. Lett. A, 174, 5-6,(1993) pp. 363-367. doi:10.1016/0375-9601(93)90191-2

[17]     Eckart, C.,The Penetration of a Potential Barrier by Electrons. Phys. Rev. 35, 1303 (1930)

[18]     C. D.J.F. (2019), Trends in Supersymmetric Quantum Mechanics. In: Kuru Ş, Negro J, Nieto L. (eds) Integrability, Supersymmetry and Coherent States. CRM Series in Mathematical Physics. Springer, Cham. doi:10.1007/978-3-030-20087-9_2.

[19]     Gangopdhayaya, A., Mallow, J., Rasinariu, C., Sypersymmetry quantum: an introduction, Second edition., Singapore; Hackensack, NJ : World Scientific, (2017)

[20]     Gendenshtein, L.E., Derivation of exact spectra of the Schrödinger equation by means of supersymmetry, JETP Lett. 38 (1983) pp. 356–359

[21]     Infeld, L., Hull, T. E. The factorization method, Rev. Mod. Phys. 23,(1951) 21–68.

[22]     Junker, G., Supersymmetry methods in quantum and statistical and solid state physics, IOP (2019).

[23]     Khare, A, Sukhatme, U. Phase equivalent potentials obtained from supersymmetry, J. Phys. A.; Math. Gen. 22, 2847, (1989).

[24]     Khare, A, and Sukhatme,U.P. Scattering Amplitudes for Supersymmetric Shape-Invariant Potentials by Operator Methods, Journal of Physics A: Mathematical and General, 21, 9, (1988), pp. L501. doi:10.1088/0305-4470/21/9/005

[25]     Khare, A, and Sukhatme,U.P.   New Shape-Invariant Potentials in Supersymmetric Quantum Mechanics, J. Phys. A: Math. Gen., 26 (1993) pp. L901-L904. doi:10.1088/0305-4470/26/18/003

[26]     Levai, G. A search for shape-invariant solvable potentials, J. Phys. A: Math. Gen, 22 (1989) pp. 689-702.

[27]     Levai, G. Solvable potentials derived from supersymmetric quantum mechanics, pp. 107-126, in: Von Geramb H.V. (eds) Quantum Inversion theory applications. Lecture Notes in Physics, V 427, Springer (1994).

[28]     Manning, M. F., Rosen, N.: Potential function of diatomic molecules. Phys. Rev. 44, 953 (1933)

[29]     Manning, M. F. (1935). Exact Solutions of the Schrödinger Equation. Phys. Rev, 48(2), 161–164. doi:10.1103/physrev.48.161

[30]     Morse, P. M. Diatomic molecules according to the wave mechanics II. Vibrational levels. Phys. Rev, 34(1), (1929) 57-64. doi:10.1103/physrev.34.57.

[31]     Nicolai, H. Supersymmetry and system spin. J. Phys. A: Math. Gen. 9 1497 (1976) Doi:10.1088/0305-4470/9/9/010.

[32]     Pöschl, G., Teller, E. Remarks on the quantum mechanics of the anharmonic oscillator. Z. Phys. 83, 143-151 (1933). https://doi.org/10.1007/BF01331132

[33]     Rosen, N., Morse, P. M. (1932). On the Vibrations of Polyatomic Molecules. Physical Review, 42(2), 210–217. doi:10.1103/physrev.42.210

[34]     Roy, B., Roy, P., & Roychoudhury, R. (1991). On Solutions of Quantum Eigenvalue Problems: A Supersymmetric Approach. Fortschritte Der Physik/Progress of Physics, 39(3), 211–258. doi:10.1002/prop.2190390304

[35]     Scarf, F. L. (1958). New Soluble Energy Band Problem. Physical Review, 112(4), 1137–1140. doi:10.1103/physrev.112.1137

[36]     Schrodinger, E., Ann. Phys., 79, 361, 1926

[37]     Schrödinger, E., 1926b, 'Quantisierung als Eigenwertproblem (Zweite Mitteilung)' Am. Phys., 79, 489

[38]     Sukumar, C. V. Supersymmetric quantum mechanics in one-dimensional systems, J. Phys. A: Math. Gen. 18, 2917 (1985).

[39]     Sukumar, C. V. Supersymmetry, factorization of the Schrödinger equation and a Hamiltonian hierarchy, J. Phys. A: Math. Gen. 18 L57 (1985).

[40]     Witten, E. Dynamical breaking of supersymmetry, Nucl. Phys. B. 188, (1981) pp 513–54. Doi: 10.1016/0550-3213(81)90006-7